 \documentclass[12pt,preprint2]{emulateapj}

\usepackage{lineno}
\usepackage{natbib}
\usepackage{amsmath}
\usepackage{graphicx}
\usepackage{wasysym}
\usepackage{txfonts}
\usepackage{xspace}
\usepackage{url}
\usepackage{textcomp}
\usepackage{multirow}
\usepackage{lipsum}

\newcommand{\vela}{Vela~X-1\xspace}

\newcommand{\inte}{\textsl{INTEGRAL}\xspace}

\newcommand{\xmm}{\textsl{XMM-Newton}\xspace}
\newcommand{\sax}{\textsl{BeppoSAX}\xspace}
\newcommand{\suz}{\textsl{Suzaku}\xspace}

\newcommand{\xte}{\textsl{RXTE}\xspace}

\newcommand{\nustar}{\textsl{NuSTAR}\xspace}

\newcommand{\snr}{S/N\xspace}

\newcommand{\msun}{\ensuremath{\text{M}_{\odot}}\xspace}

\newcommand{\redchi}{\ensuremath{\chi^{2}_\text{red}}\xspace}

\newcommand{\feka}{\ensuremath{\mathrm{Fe}~\mathrm{K}\alpha}\xspace}
\newcommand{\fekb}{\ensuremath{\mathrm{Fe}~\mathrm{K}\beta}\xspace}
\newcommand{\nh}{\ensuremath{{N}_\mathrm{H}}\xspace}

\newcommand{\nhone}{\ensuremath{{N}_{\mathrm{H},1}}\xspace}
\newcommand{\nhtwo}{\ensuremath{{N}_{\mathrm{H},2}}\xspace}

\shorttitle{Luminosity dependent cyclotron line energy in Vela~X-1}
\shortauthors{F\"urst et al.}

\begin{document}



\title{\textsl{NuSTAR} discovery of a luminosity dependent cyclotron line energy in Vela~X-1}


\author{Felix F\"urst\altaffilmark{1}}
\author{Katja Pottschmidt\altaffilmark{2,3}}
\author{J\"orn Wilms\altaffilmark{4}}
\author{John A. Tomsick\altaffilmark{5}}
\author{Matteo Bachetti\altaffilmark{6,7}}
\author{Steven E. Boggs\altaffilmark{5}}
\author{Finn E. Christensen\altaffilmark{8}}
\author{William W. Craig\altaffilmark{5,9}}
\author{Brian W. Grefenstette\altaffilmark{1}}
\author{Charles J. Hailey\altaffilmark{10}}
\author{Fiona Harrison\altaffilmark{1}}
\author{Kristin K. Madsen\altaffilmark{1}}
\author{Jon M. Miller\altaffilmark{11}}
\author{Daniel Stern\altaffilmark{12}}
\author{Dominic J. Walton\altaffilmark{1}}
\author{William Zhang\altaffilmark{13}}

\altaffiltext{1}{Cahill Center for Astronomy and Astrophysics, California Institute of Technology, Pasadena, CA 91125}
\altaffiltext{2}{Center for Space Science and Technology, University of Maryland
Baltimore County, Baltimore, MD 21250, USA}
\altaffiltext{3}{CRESST and NASA Goddard Space Flight Center, Astrophysics Science
Division, Code 661, Greenbelt, MD 20771, USA}
\altaffiltext{4}{Dr. Karl-Remeis-Sternwarte and ECAP, Sternwartstr. 7, 96049 Bamberg, Germany}
\altaffiltext{5}{Space Sciences Laboratory, University of California, Berkeley, CA 94720, USA}
\altaffiltext{6}{Universit\'e de Toulouse; UPS-OMP; IRAP; Toulouse, France}
\altaffiltext{7}{CNRS; Institut de Recherche en Astrophysique et Plan\'etologie; 9 Av. colonel Roche, BP 44346, 31028 Toulouse cedex 4, France}
\altaffiltext{8}{DTU Space, National Space Institute, Technical University of Denmark, Elektrovej 327, 2800 Lyngby, Denmark}
\altaffiltext{9}{Lawrence Livermore National Laboratory, Livermore, CA 94550, USA}
\altaffiltext{10}{Columbia Astrophysics Laboratory, Columbia University, New York, NY 10027, USA}
\altaffiltext{11}{Department of Astronomy, The University of Michigan, Ann Arbor, MI 48109, USA}
\altaffiltext{12}{Jet Propulsion Laboratory, California Institute of Technology, Pasadena, CA 91109, USA}
\altaffiltext{13}{NASA Goddard Space Flight Center, Astrophysics Science
Division, Code 662, Greenbelt, MD 20771, USA}


\begin{abstract}
We present \nustar observations of \vela, a persistent, yet highly
variable, neutron star high-mass X-ray binary (HMXB). 
Two observations were taken at similar orbital phases but separated by nearly a year.
They show
very different 3--79 keV flux levels as well as strong variability during each observation, covering almost one order of magnitude in flux.
 These observations allow, for the first time ever, investigations on kilo-second time-scales
of how  the centroid energies of cyclotron resonant scattering features (CRSFs) depend on flux for a persistent HMXB.  We find
that the line energy of the harmonic CRSF is correlated with flux,
as expected in the sub-critical accretion regime. We argue that
\vela has a very narrow accretion column with a radius of around 0.4\,km that
sustains a Coulomb interaction dominated shock at the observed luminosities
of $L_\text{x}\sim3\times10^{36}$\,erg\,s$^{-1}$.
Besides the prominent harmonic line at 55\,keV the fundamental line around 25\,keV is clearly detected. We find that the strengths of the two CRSFs are anti-correlated, which we explain by photon spawning. This anti-correlation is a possible explanation for the debate about the existence of the fundamental line. The ratio
of the line energies is variable with time and deviates significantly
from 2.0, also a possible consequence of photon spawning, which changes the shape of the line.
During the second observation, \vela showed a short
off-state in which the power-law softened and a 
cut-off was no longer measurable. It is likely that the source switched to a different accretion regime at these low mass accretion rates, explaining the drastic change in spectral shape.

\end{abstract}


\keywords{stars: neutron, X-rays: individual (Vela X-1), X-rays: binaries, radiation: dynamics}



\section{Introduction}
\vela is one of the most famous wind-accreting neutron star X-ray binaries. The neutron star, rotating with a period of about 283\,s around its spin axis, is in an 8.9\,d orbit around the B0.5Ib super-giant HD\,77523 \citep{hiltner72a}. With an orbital separation of only $1.7\,\text{R}_\star$ \citep{joss84a,quaintrell03a}, the neutron star is deeply embedded in the stellar wind, accreting on the order of $10^{-9}$\,\msun\,yr$^{-1}$ \citep[and references therein]{fuerst10a}. The distance to the system is $1.9\pm0.2$\,kpc \citep{sadakane85a}, making it one of the brightest persistent X-ray binaries in the sky, despite a mean luminosity of only $5\times10^{36}$\,erg\,s$^{-1}$ \citep{fuerst10a}.  The neutron star seems to be significantly more massive than the canonical mass of 1.4\,\msun, with a lower limit of about $M_\star=1.8\,\msun$ \citep{vankerkwijk95a, quaintrell03a}.  Its flux is strongly variable, changing from barely detectable during off-states to up to 7\,Crab during giant flares \citep[in the 20--40\,keV  energy band, ][]{kreykenbohm08a, staubert04a}. This volatile behavior is a consequence of direct wind accretion, in which the material couples to the magnetic field before a stable accretion disk can form which then would mediate the accretion flow.

The wind of the super-giant is highly structured, or clumpy, which is a result of line-driven instability and the influence of the neutron star \citep[see, among others, ][and references therein]{owocki88a, dessart05a,  ducci09a, fuerst10a,  oskinova12a}. These clumps manifest themselves in two ways in the X-rays: as partially covering, variable absorption, and through a highly variable accretion rate. From the duration and brightness of flares, \citet{fuerst10a}, using \inte data taken between 2003 and 2006, estimated that typical clumps have masses of $M\approx5\times10^{19}$\,g and radii of $r\approx2\times10^{10}$\,cm, assuming a spherical clump geometry and homogeneous density. Similar values have also been found by \citet{odaka13a} using \suz data and by \citet{martinez13a} using \xmm data. The absorption column density of such clumps is around $2\times10^{22}$\,cm$^{-2}$, explaining naturally the \nh variations seen at early orbital phases by clumps moving through our line-of-sight \citep[see, e.g., ][]{haberl90a,pan94a}.

Cyclotron resonant scattering features (CRSFs, cyclotron lines for short) are an important feature of the X-ray spectra of accreting neutron stars. They are produced in strong magnetic fields, where electrons are quantized onto Landau-levels \citep[and references therein]{schoenherr07a}. The energy of these levels is a strict function of the magnetic field strength. Photons with energies close to the Landau-levels are removed from the observed X-ray spectrum by scattering off these electrons. CRSFs appear as broad absorption lines due to Doppler broadening, and their energy is the only way to directly measure the magnetic field close to the surface of neutron stars. \vela shows a prominent harmonic line at 55\,keV and a weaker fundamental line at 25\,keV \citep{kendziorra92a, kretschmar96a, kretschmar97a, kreykenbohm02a, maitra13a}. The magnetic field strength can be calculated using
\begin{equation}
B = \frac{E_n}{n\times11.57\,\text{keV}} \times 10^{12}\,\text{G}\quad,
\end{equation} 
where $E_n$ is the energy of the fundamental ($n=1$) or harmonic line ($n=2$) line in keV. 

The detection of fundamental lines is complicated by the fact that in hard sources, with strong harmonic lines, its shape can be drastically altered to a point where the line is almost filled by spawned photons. Photon-spawning occurs when an electron remains in an excited Landau state after resonantly scattering, emitting another photon of similar energy when de-exciting. These photons can escape the line-forming plasma, especially when their energy is at the wings of the fundamental line. Theoretical calculations by \citet{schoenherr07a} show that with increasing number of harmonic lines, the fundamental line gets shallower and can become undetectable in the observed X-ray spectrum. 

The centroid line energy of the CRSF is often found to vary with X-ray flux. This behavior was first observed with high significance in the transient source V\,0332+53, in which the line energy decreases with increasing flux \citep[see Figure~\ref{fig:lx2ecyc}]{tsygankov10a, tsygankov06a, mowlavi06a}. On the other hand, a positive correlation was found in Her~X-1 \citep{staubert07a} and GX~304$-$1 \citep{yamamoto11a, klochkov12a}. In all these measurements, care has to be taken that the underlying continuum does not influence the line energy, as demonstrated for 4U~0115+634, where a anti-correlation was found using the NPEX continuum \citep{nakajima06a}, but not when using a simpler cutoff-powerlaw \citep{mueller13a}.

\citet{klochkov11a} performed a detailed study of the correlation between line energy and flux, confirming the behavior of Her~X-1 and V~0332+53 on shorter time-scales. 
They also found conclusive evidence that the power-law photon-index is changing with flux, and that it shows the opposite behavior to the line energy, i.e., that it decreases with flux in sources with a positive line-energy to flux correlation, and vice versa.

Both phenomena, the changing energy and the changing photon-index can be understood in terms of a change in altitude of the X-ray production region as function of flux, an idea put forward by \citet{basko76a} and discussed by \citet{klochkov11a}, \citet{staubert07a}, and \citet{mowlavi06a}, among others. With changing altitude, the local magnetic field changes, explaining the change in energy. At the same time, the local plasma temperature changes, which can result in a change in the observed photon-index. Depending on the overall luminosity, the correlation can switch from negative to positive, as described below, following the theory described in \citet{becker12a}.
	

At typical luminosities of \vela ($L_\text{x} \leq 10^{37}$\,erg\,s$^{-1}$), the matter is expected to be stopped very close to, or at, the stellar surface, so that the location of the line forming region is independent of the X-ray flux \citep{becker12a}. At higher luminosities, a shock dominated by Coulomb interactions can form, stopping the matter above the stellar surface.  The ram pressure of the in-falling material increases with higher accretion rate and pushes the shock region down into regions of higher magnetic field strength. This leads to a positive correlation of the line energy with flux and a hardening of the power-law component \citep[as suggested by][]{staubert07a}.
If the local Eddington luminosity is exceeded, at the so-called critical luminosity, $L_\text{crit} \approx 3\times10^{37}$\,erg\,s$^{-1}$ for typical neutron star parameters, a radiation-dominated shock forms. The shock height is then positively correlated with the accretion rate as higher X-ray luminosity leads to a higher radiation pressure. The results is an observed an anti-correlation of the CRSF energy and correlation of the photon-index with flux.

The line behavior has not been investigated for \vela, since no existing observations cover the necessary large range of luminosities with high spectral resolution. However, in a recent study using \suz, \citet{odaka13a} found that the photon index $\Gamma$ is anti-correlated with the flux. This correlation indicates a sub-critical luminosity and would put \vela in the same group of sources as Her~X-1 and GX~304$-$1. In the study by \citet{odaka13a} the CRSF parameters were fixed in time-resolved spectroscopy and could not be studied as a function of luminosity. 



In this paper we perform a spectral analysis on short-time scales for two separate data-sets taken by the \textsl{Nuclear Spectroscopic Telescope Array} \citep[\nustar; ][]{harrison13a} about 1\,yr apart. In \S\ref{sec:data} we describe the data, their calibration and the software used. \S\ref{sec:lc} discusses the lightcurve and variability of the hardness with time. \S\ref{sec:phasavg} presents the analysis of the time-averaged spectrum. In \S\ref{sec:tresspec} we perform time-resolved spectroscopy on pulse-to-pulse as well as on longer time-scales. In the \S\ref{sec:disc} we discuss our results and give a physical interpretation of the parameters. In \S\ref{sec:summ} we summarize our findings and give an outlook to future work.

\section{Observations and data reduction}
\label{sec:data}
\nustar consists of two independent gracing incidence telescopes, focusing X-rays between 3.0--79\,keV on corresponding focal planes consisting of cadmium zinc telluride (CZT) pixel detectors. \nustar provides unprecedented sensitivity and high spectral resolution at hard X-rays, ideally suited to study CRSFs in \vela. The two focal planes are referred to as focal plane module (FPM) A and B.

\nustar observed \vela twice, once early in the mission as a calibration target for $\sim$11\,ks and later as a science target for $\sim$42\,ks. We refer to these data-sets as observation I and II, respectively. Table~\ref{tab:log} gives details about the observations, including the orbital phase, derived from the ephemeris by \citet{kreykenbohm08a}. While the orbital phase was very similar in both observations, \vela was  a factor of about 3.5 brighter during observation I. 
Despite the shorter observation time of observation I, it provides similar \snr for the phase-averaged data due to the higher flux.

\begin{deluxetable}{ccccc}
\tablecolumns{5}
\tablewidth{0pc}
\tablecaption{Observation log\label{tab:log}}
\tablehead{\colhead{ ObsID} &\colhead{time start} & \colhead{time stop} & \colhead{exposure} & \colhead{orbital}  \\
\colhead{} & \colhead{MJD [d]} & \colhead{ MJD [d]} & \colhead{(FPMA/B) [ks]} &  \colhead{phase}}
\startdata
   10002007 & 56117.6339 & 56117.9441 & 10.82/10.96 & 0.655--0.690 \\
 30002007 & 56404.5275 & 56405.5559 & 41.67/41.92 & 0.659--0.773
\enddata
\end{deluxetable}

We used the latest NUSTARDAS pipeline v1.2.0 and HEASOFT v6.13 to extract spectra and lightcurves. The exposure times given in Table~\ref{tab:log} are the final exposure times after filtering for occultations, SAA passages and taking the detector dead-time into account. We  barycentered the event times with the FTOOL \texttt{barycen} using the DE-200 solar system ephemeris and corrected for the binary orbit, using the ephemeris by \citet{kreykenbohm08a}. The data were analyzed using the Interactive Spectral Interpretation System (ISIS) v1.6.2 \citep{houck00a}. 

We extracted source spectra from a region with $150''$ radius around \vela's FK5 coordinates, separately for FPMA and FPMB. As \vela is a very bright source, source photons illuminate the whole focal plane. To minimize their influence on the background estimation we extracted background spectra from a $80''$ radius region as far away from the source as possible. Since the background changes over the field-of-view \citep{wik13a}, systematic uncertainties are formally introduced by this method. \vela, however, is about a factor 5 brighter than the background even at the highest energies, such that the effect of residual uncertainties is negligible.


Both spectra were rebinned within ISIS to a minimal \snr of 32, 24, 20, 8, and 2 adding at least 2, 4, 8, 12, and 18 channels for energies between 3.0--10, 10--28, 28--42, 42--54, and 54--79\,keV, respectively. This rebinning allows us to use $\chi^2$ statistics for the fit. Uncertainties are given at the 90\% confidence level ($\Delta \chi^2=2.7$ for one parameter of interest), unless otherwise noted.

\section{Lightcurves and pulse profiles}
\label{sec:lc}
 The pulse-period of \vela varies in a random-walk like manner \citep[and references therein]{dekool93a}, as is typical for a wind-accretor \citep{bildsten97a}. 
We use the pulse period as the minimal integration time for our time-resolved spectroscopy, we therefore need to measure its value accurately. To do so, we  used epoch-folding \citep{leahy87a} on the 3.0--79\,keV lightcurve of FPMA with 1\,s time resolution. We find the period during observation II to be
\begin{equation}
P = 283.4290 \pm 0.0006\,\text{s}\quad.
\end{equation}
The uncertainty is estimated using a Monte Carlo simulation technique as described by \citet{larsson96a} with 2000 runs. 
 The period is also consistent with the epoch folding results from observation I, with an uncertainty of $\Delta P = 0.0024$\,s.


\begin{figure*}
\includegraphics[width=0.96\textwidth]{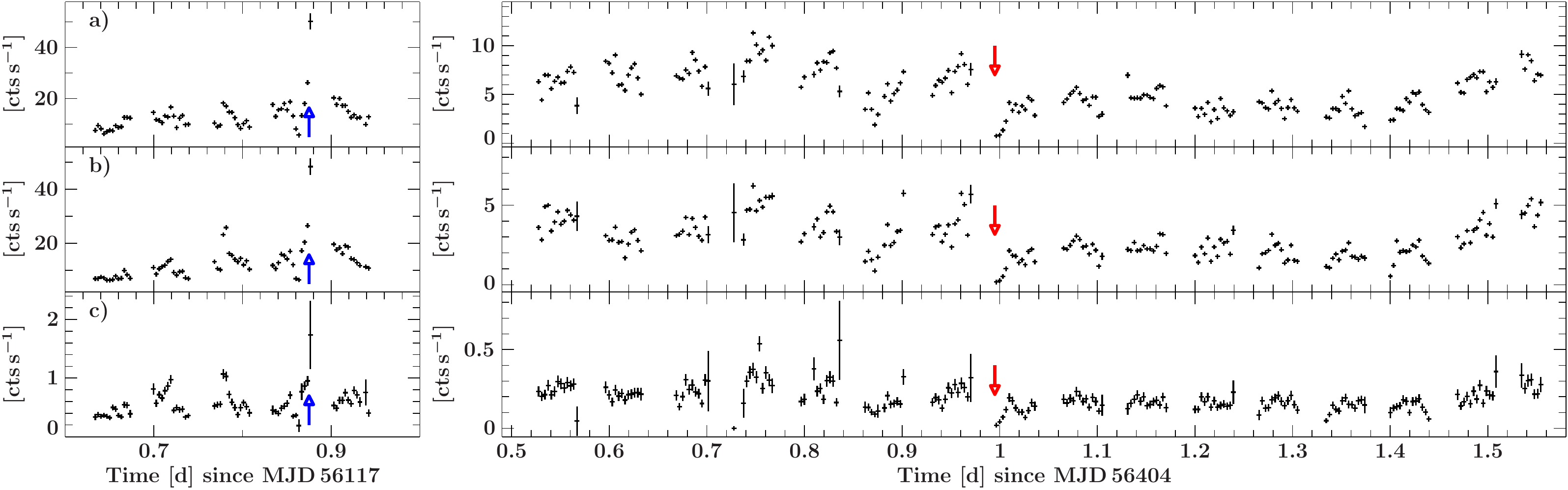} 
\caption{Lightcurves for observation I (left) and observation II (right) with a time resolution of 283.43\,s (the pulse period) in different energy bands: \textit{a)} 3--5\,keV, \textit{b)} 20--30\,keV, and \textit{c)} 40--79\,keV. Note that the $y$-scale is different between the two observations. The times of a major flare and an off-state are indicated by blue upwards-pointing arrows  and the red downwards-pointing arrows, respectively. }
\label{fig:lcall}
\end{figure*}

\vela is known to be highly variable on all time-scales, as is typical for a wind-accretor. Even within only a few pulse-periods, the flux can change by a factor of 5--7 \citep{kreykenbohm08a}. This strong variability is also seen in the \nustar lightcurves (Figure~\ref{fig:lcall}). In observation I, a large flare was detected, indicated in Figure~\ref{fig:lcall}. Here, the dead-time corrected 3--79\,keV count-rate increased to  $\approx$600\,cts\,s$^{-1}$, from an average of $\approx$150\,cts\,s$^{-1}$ during the whole observation. Observation II was  fainter overall with an average 3--79\,keV count-rate of only $\approx$48\,cts\,s$^{-1}$ and it did not show huge flares. Instead, an off-state was detected, highlighted in Figure~\ref{fig:lcall}. In this off-state, the count-rate dropped to 3--4\,cts\,s$^{-1}$. 

\begin{figure*}
\includegraphics[width=0.96\textwidth]{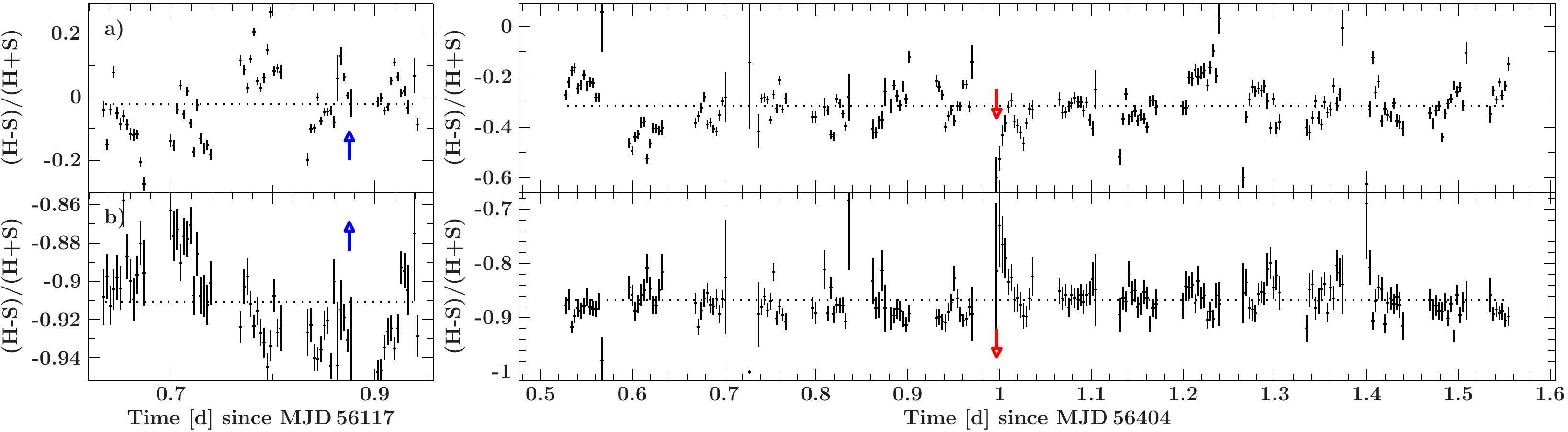} 
\caption{Hardness-ratios for observation I (left) and observation II (right) with a time resolution of 283.43\,s between \textit{a)} $S=3$--5\,keV and $H=20$--30\,keV and \textit{b)} $S=20$--30\,keV and  $H=40$--79\,keV. The dashed lines indicate the average value. Note that the $y$-scale is different between the two observations. The times of a major flare and an off-state are indicated by blue upwards-pointing arrows  and the red downwards-pointing arrows, respectively. }
\label{fig:hrall}
\end{figure*} 

Flux variability is often associated with a change in the spectrum as well. In Figure~\ref{fig:hrall} we show the hardness-ratios HR, calculated as $\text{HR}=(H-S)/(H+S)$, between three energy bands (a low-energy ratio in the top panel, and a high-energy ratio in the bottom panel). The flare in observation I does not seem to be associated with a strong spectral change, especially for the ratio between $S=3$--5\,keV and $H=20$--30\,keV. This behavior is similar to many of the flares studied by \citet{staubert04a}, \citet{kreykenbohm08a} and \citet{odaka13a}.

During the off-state in observation II, strong changes in both hardness ratios are evident, with the spectrum becoming softer in the low-energy ratio and harder in the high-energy ratio, i.e., the spectrum becomes flatter overall. A detailed analysis of off-states in \suz data is presented by \citet{doroshenko11a}, confirming this overall behavior.


The strong changes of both flux and hardness necessitate an analysis of the spectral variability on short time-scales. We therefore extracted one spectrum per pulse. However, the \snr in each individual spectrum is not high enough to fit all parameters in a typical absorbed power-law model with a high-energy cutoff. We therefore first fit the phase-averaged spectra of both observations separately in \S\ref{sec:phasavg} and use these parameters for the pulse-to-pulse fits in \S\ref{sec:tresspec}.


\section{Time-averaged spectroscopy}
\label{sec:phasavg}

To model the time-averaged spectrum, we use a power-law continuum with a {Fermi-Dirac cut-off} \citep[\texttt{FDcut, }][]{tanaka86a}. This model is  often applied to \vela \citep[see, e.g., ][]{kreykenbohm08a}.  A simple \texttt{cutoffpl} does not fit the data, and the \texttt{NPEX} model \citep{makishima99a} can result in unrealistic CRSF line parameters \citep{mueller13a}. To confirm our results, we also applied the \texttt{NPEX} model, as described in \S\ref{sec:tresspec}.
The \texttt{FDcut}-model is characterized by the photon index $\Gamma$, the cut-off energy $E_\text{cut}$ and the folding energy $E_\text{fold}$ in the following form:
\begin{equation}
F(E) \propto E^{-\Gamma} \times \left(1+\exp\left(\frac{E-E_\text{cut}}{E_\text{fold}}\right)\right)^{-1} \quad .
\end{equation}

Even though the two telescopes of \nustar are almost identical, small differences in the measured flux are possible, with the difference being below 5\%. To allow for this cross-calibration difference, we include a constant in the model, $\text{C}_\text{FPMB}$, which gives the relative normalization of FPMB with respect to FPMA.

The neutron star is embedded deeply in the stellar wind and an intrinsic absorption column of the order of $10^{23}$\,cm$^{-2}$ is often observed at the late  orbital phases where both \nustar observations took place ($\sim 0.6$; see Table~\ref{tab:log}). We describe this absorption column using an updated version of the \texttt{tbabs} model \citep{wilms00a} and the corresponding abundances and cross-sections \citep{verner96a}. The absorbing material is not smooth, but structured or clumpy \citep[see, e.g., ][]{fuerst10a}. To take this structure into account, we use a partial covering model with two absorption columns, \nhone and \nhtwo, and a covering fraction $0<\text{CF}<1$.

\vela also shows fluorescence lines in its soft X-ray spectrum, associated with \feka and \fekb. Their energies are  usually close to that of neutral iron \citep{watanabe06a}. We modeled these lines with two independent Gaussians around 6.4\,keV and 7.1\,keV. For neutral iron we would expect that the flux of \fekb is about 0.13 the flux \feka \citep{palmeri03a}. However, because the overlapping Fe~K-edge at 7.1\,keV, the \fekb flux is difficult to constrain with \nustar and can be artificially higher than expected from theoretical calculations.

To describe the cyclotron lines we use a multiplicative absorption line model, $\text{CRSF}(E)=\exp\left(-\tau(E)\right)$, with a Gaussian optical depth profile (\texttt{gabs} in XSPEC):
\begin{equation}
\tau(E) = \frac{d}{\sigma\sqrt{2\pi}}\exp\left(-\frac12\left(\frac{E-E_\text{cyc}}{\sigma}\right)^2\right)\quad,
\end{equation}
with line depth $d$ and width $\sigma$. We use the subscripts ``CRSF,F'' and ``CRSF,H'' to denote parameters for the fundamental and the harmonic line, respectively.

When we apply this model without allowing for a fundamental line, the overall continuum shape is well described, but the obtained $\chi^2$ values are unacceptable ($\chi^2/$d.o.f. = 791.4/559 for observation I and  897.1/553 for observation II). Adding the fundamental line does results in almost the same quality of fit for observation I but improves it dramatically for observation II ($\chi^2/$d.o.f. = 789.7/557 and  614.2/551, respectively). 
Since the fundamental line is relatively weak, and its width unconstrained, we required its width $\sigma_\text{CRSF,F}$ to be half the width of the harmonic line, $\sigma_\text{CRSF,H}$ in both observations, as indicated by  the best-fit of observation II. A narrow fundamental line  prevents it from becoming so strong as to unphysically influence the other model parameters. \citet{kreykenbohm02a} also found that the fundamental line is consistently narrower in pulse-resolved analysis.



Residuals remain around 10\,keV, which can be very well described with a broad Gaussian absorption line. Its physical origin is unexplained, but it has been observed in \vela before \citep[e.g., ][]{labarbera03a}, as well as in many other CRSF sources \citep{coburn02a}, always at $\sim$10\,keV.
The current version of the \nustar responses show higher uncertainties around the tungsten edge at $\sim$10\,keV. These response features are, however, clearly weaker and have a different shape than the residuals described here \citep{fuerst13a}.
The 10\,keV feature is more prominent in observation I and the final fit gives values of $\chi^2/$d.o.f. = 587.5/554 and 570.8/548 for the two observations, respectively.

This final model can be summarized as
\begin{align}
I(E) = & \left(\text{CF}\times\nhone  + \left(1-\text{CF}\right) \times \nhtwo\right) \times \notag\\
& \times\Bigl(\text{FDcut} \times\text{CRSF(F)}\times\text{CRSF(H)}+ \notag\\
& +\text{10\,keV-line}+\feka+\fekb\Bigr)\quad.
\end{align}

\begin{figure*}
\includegraphics[width=0.96\columnwidth]{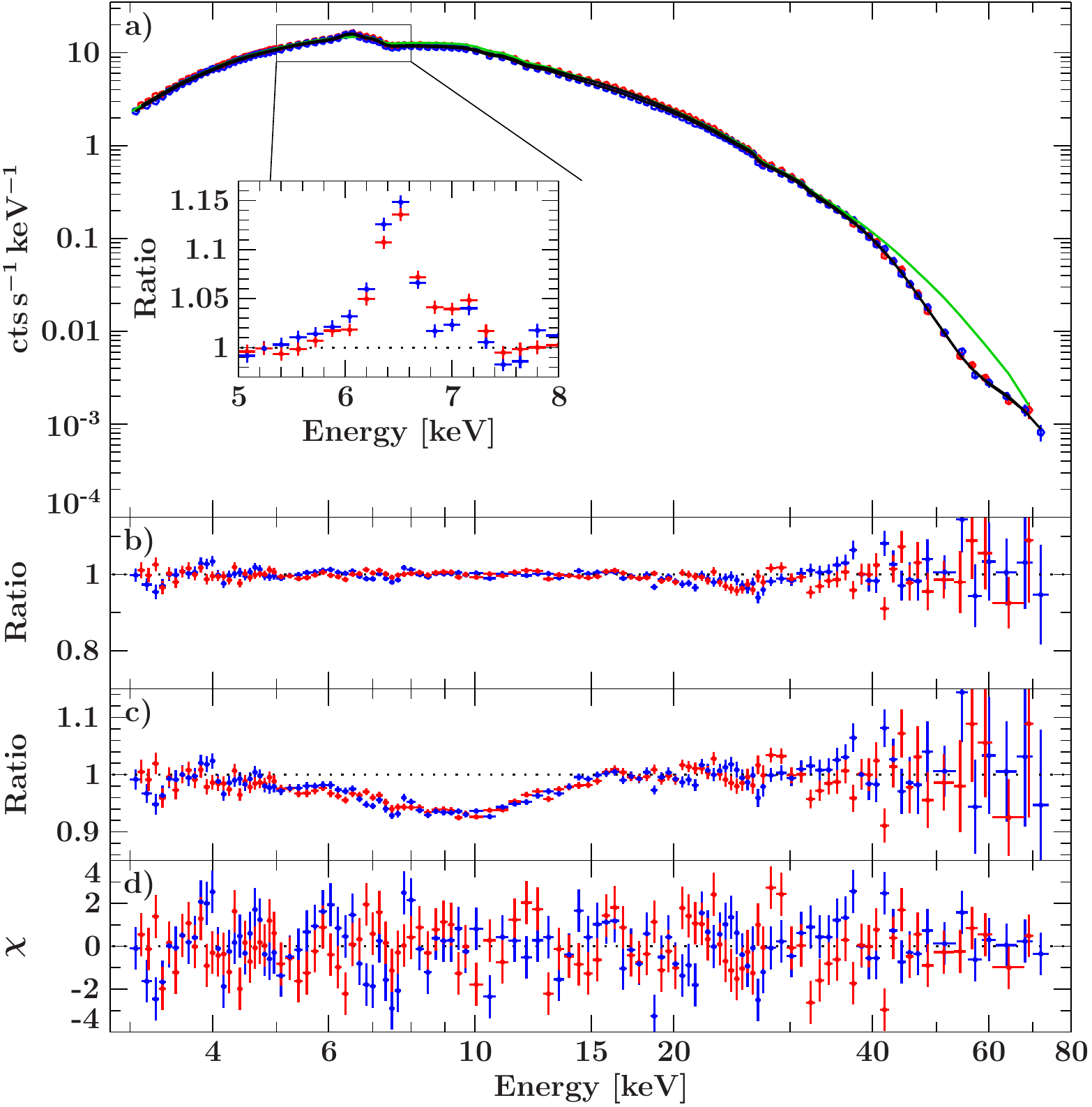}\hfill 
\includegraphics[width=0.96\columnwidth]{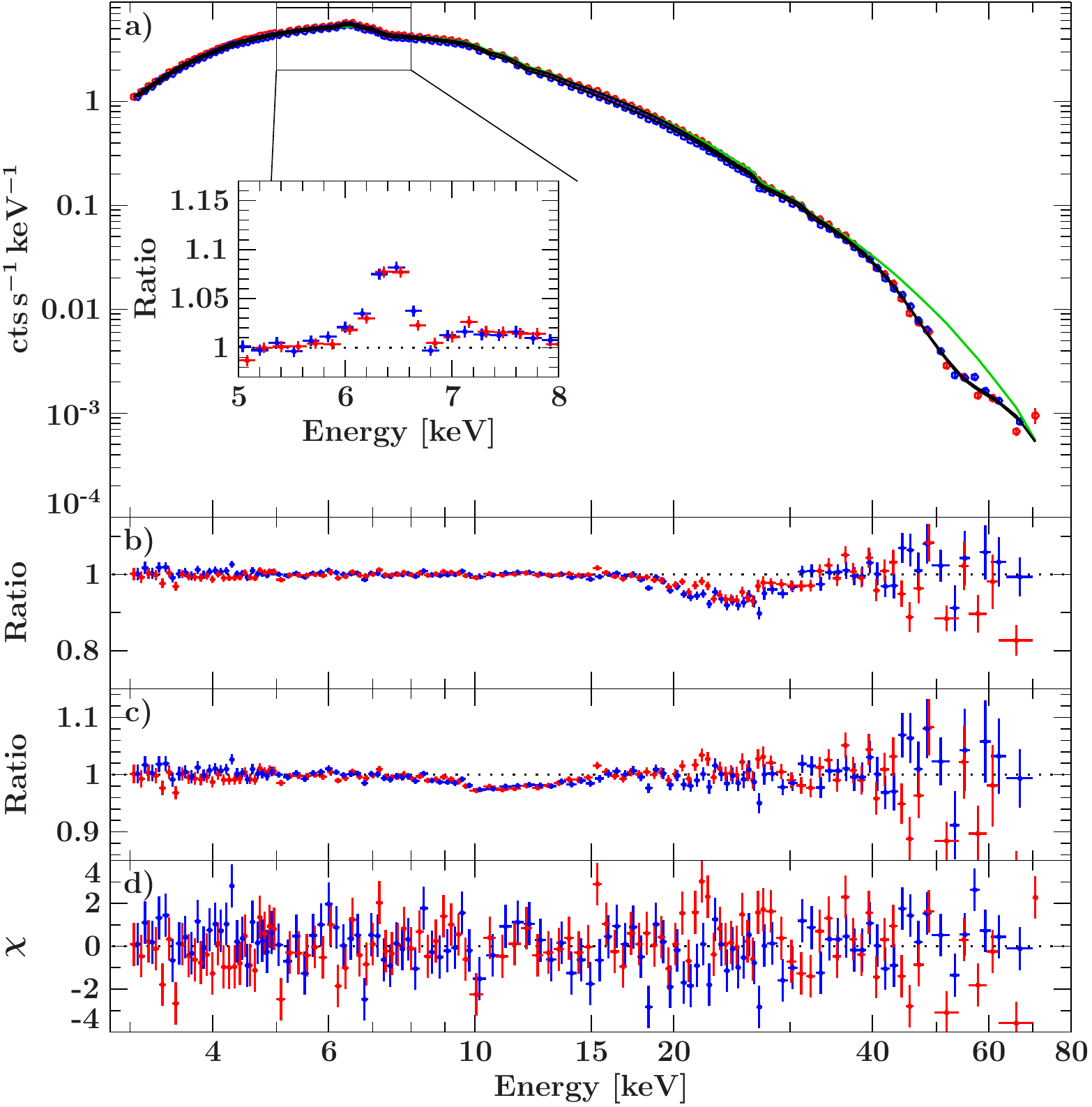} 
\caption{\textit{a)} Spectrum and best-fit model for observation I (left) and observation II (right), FPMA data in red and FPMB data in blue. The best fit model is shown in black, the continuum without the absorption lines in green.  The insets show the respective residuals in the iron line region without the two Gaussian lines. \textit{b)} Residuals without the fundamental CRSF line. \textit{c)} Residuals without the 10\,keV-feature. \textit{d)} Residuals to the best-fit in units of $\sigma$.}
\label{fig:spec_all_I}
\end{figure*}

The data and best-fit model for both observations are shown in Figure~\ref{fig:spec_all_I}. In the residual panels \textit{b} and \textit{c} in both figures the fundamental line and the 10\,keV feature are turned off, respectively, to show their contribution to the overall continuum shape. It is clear that both features are variable, and that the fundamental line is almost invisible in observation I. Table~\ref{tab:bestfit_all} lists the best-fit parameters for both observations. Besides the variability of the cyclotron lines, the photon index $\Gamma$  also changes significantly between the two observations, while the cutoff and folding energy are almost constant.
The second absorber \nhtwo could not be constrained in observation II, so we fixed it to 0.

\begin{deluxetable}{rll}
\tablewidth{0pc}
\tablecaption{Fit parameters for the best-fit time-averaged model.\label{tab:bestfit_all}}
\tablehead{\colhead{Parameter}  & \colhead{ Obs. I} & \colhead{Obs. II}}
\startdata
 $ CF$ & $0.823^{+0.022}_{-0.053}$ & $0.868\pm0.007$ \\
 $ N_\text{H,1} [10^{22}\text{\,cm}^{-2}]$ & $34.2^{+5.7}_{-2.1}$ & $26.0^{+0.9}_{-1.0}$ \\
 $ N_\text{H,2} [10^{22}\text{\,cm}^{-2}]$ & $0.9^{+4.2}_{-0.9}$ & --- \\
 $ \Gamma$ & $0.88^{+0.05}_{-0.06}$ & $1.28^{+0.04}_{-0.05}$ \\
 $ E_\text{cut} [\text{keV}]$ & $19.3^{+2.5}_{-3.8}$ & $22.3^{+2.8}_{-3.9}$ \\
 $ E_\text{fold} [\text{keV}]$ & $10.5^{+0.9}_{-0.5}$ & $13.3^{+1.0}_{-0.8}$ \\
 $ E_\text{CRSF,H} [\text{keV}]$ & $55.4^{+1.3}_{-1.0}$ & $53.2^{+0.9}_{-0.8}$ \\
 $ \sigma_\text{CRSF,H} [\text{keV}]$ & $8.0^{+1.3}_{-1.0}$ & $6.8^{+0.9}_{-0.7}$ \\
 $ d_\text{CSRF,H} [\text{keV}]$ & $20^{+8}_{-5}$ & $14.0^{+3.2}_{-2.3}$ \\
 $ E_\text{CRSF,F} [\text{keV}]$ & $25.1^{+2.0}_{-2.4}$ & $24.8\pm0.6$ \\
 $ d_\text{CSRF,F} [\text{keV}]$ & $0.24^{+0.20}_{-0.13}$ & $0.61^{+0.18}_{-0.12}$ \\
 $ A(\text{Fe\,K}\alpha)^\dagger$ & $0.038^{+0.263}_{-0.019}$ & $\left(0.93^{+0.09}_{-0.08}\right)\times10^{-3}$ \\
 $ E(\text{Fe\,K}\alpha) [\text{keV}]$ & $6.474\pm0.010$ & $6.4393^{+0.0008}_{-0.0393}$ \\
 $ \sigma(\text{Fe\,K}\alpha)[\text{keV}]$ & $0.065^{+0.030}_{-0.065}$ & $\le0.07$ \\
 $ A(\text{Fe\,K}\beta)$\tablenotemark{$\dagger$} & $0.008^{+0.068}_{-0.008}$ & $\left(3.3^{+1.3}_{-1.0}\right)\times10^{-4}$ \\
 $ E(\text{Fe\,K}\beta) [\text{keV}]$ & $7.084^{+0.077}_{-0.008}$ & $7.39\pm0.12$ \\
 $ \sigma(\text{Fe\,K}\beta)[\text{keV}]$ & $0.00022^{+0.07167}_{-0.00022}$ & $0.30^{+0.16}_{-0.12}$ \\
 $ A(\text{10\,keV})$ & $-0.19^{+0.14}_{-1.82}$ & $\left(-1.3^{+0.5}_{-0.9}\right)\times10^{-3}$ \\
 $ E(\text{10\,keV}) [\text{keV}]$ & $8.6^{+0.5}_{-0.7}$ & $10.9\pm0.4$ \\
 $ \sigma(\text{10\,keV})[\text{keV}]$ & $2.8^{+0.6}_{-0.5}$ & $1.9^{+0.6}_{-0.5}$ \\
 $ \mathcal{F}_\text{3--79\,keV}$\tablenotemark{$\star$} & $11.22^{+0.10}_{-0.08}$ & $3.206\pm0.016$ \\
 $ \text{C}_\text{FPMB}$ & $1.0491\pm0.0020$ & $1.0259\pm0.0018$ \\
 \tableline
 $\chi^2/\text{d.o.f.}$   & 616.63/565& 589.12/560\\$\chi^2_\text{red}$   & 1.091& 1.052
\enddata
\tablenotetext{$^\dagger$} {in ph\,s$^{-1}$\,cm$^{-2}$} 
\tablenotetext{$^\star$}{in keV\,s$^{-1}$\,cm$^{-2}$}
\end{deluxetable}

%
%

\section{Time-resolved spectroscopy}
\label{sec:tresspec}
Having established a good description of the time-averaged spectrum, we use this model to study the variability of the spectral parameters on a pulse-to-pulse basis.
Thanks to the high sensitivity of \nustar, it is possible to obtain a high quality spectrum for each individual pulse, i.e., rotation of the neutron star. 
Higher time resolution is not sensible, as the spectrum changes strongly with pulse phase \citep[see, e.g., ][]{kreykenbohm02a, labarbera03a, maitra13a}. 
This approach is similar to the analysis of \citet{fuerst11b} for GX~301$-$2 using \xmm, but thanks to \nustar's broad band coverage changes in the continuum and the cyclotron line can also be studied.

We applied the time-averaged model, but due to the reduced \snr in the pulse-to-pulse spectra we had to fix some parameters to their respective time-averaged values.
We only let the  covering fraction CF, one absorption column \nhone, the photon index $\Gamma$, and the normalization of all lines or line-like features (fundamental CRSF, harmonic CRSF, 10\,keV feature, as well as \feka and \fekb) vary. We chose the photon index $\Gamma$ as the variable continuum parameter, as it shows the strongest variation between observation I and II. With this approach we obtained very good \redchi values for all spectra, with an average $\redchi\approx1.05$. For the observation I fits, \nhtwo was fixed to $9\times10^{21}\,\text{cm}^{-2}$, while for observation II it was set to 0, the respective values of the time-averaged models.

\begin{figure*}
\includegraphics[width=0.96\textwidth]{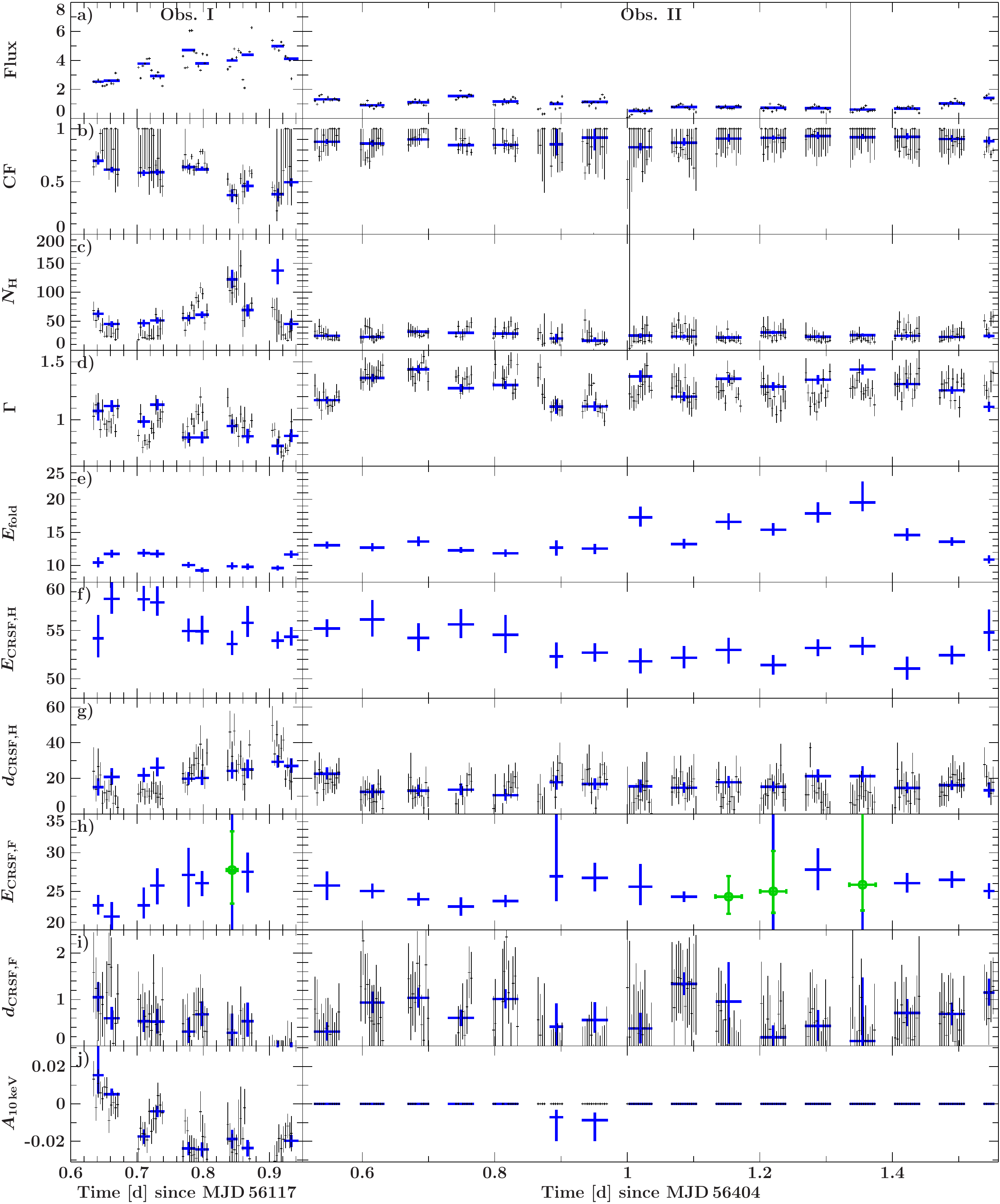} 
\caption{Results of the pulse-to-pulse (black) and ks-integrated (blue) spectral fits for both observations (observation I in the left column, observation II in the right column). \textit{a)} Unabsorbed flux in the 20--40\,keV energy band in keV\,s$^{-1}$\,cm\,$^{-2}$, \textit{b)} covering fraction, \textit{c)} absorption column in $10^{22}$\,cm$^{-2}$, \textit{d)} photon index $\Gamma$, \textit{e)} folding energy $E_\text{fold}$ in keV, \textit{f)} centroid energy of the harmonic CRSF in keV; \textit{g)} depth of the harmonic CRSF in keV; \textit{h)} centroid energy of the fundamental CRSF in keV, green values are results with the depth frozen to its best fit value, see text for details; \textit{i)}  depth of the fundamental CRSF in keV; and \textit{j)} normalization of the 10\,keV-feature in ph\,s$^{-1}$\,cm$^{-2}$. }
\label{fig:p2pres_all}
\end{figure*} 

Figure~\ref{fig:p2pres_all} show the measured variation of the model parameters as a function of time. 
All parameters show variability with time, most prominently \nhone, $\Gamma$, and CF. However, the typical time-scale for significant changes seems to be longer than single pulses.
We use the fact that consecutive data often show similar spectral parameters to add them up in order to obtain better \snr. During observation I, we added half of each $\sim$90\,min satellite orbit, yielding $\sim$1.1\,ks integration time per spectrum. We removed the high \nh data points around MJD~56117.85. Observation II was less bright, so we integrated mostly over one satellite orbit, i.e., about 2.4\,ks, cutting out the dip around MJD~56405.0 and the phase of dramatic change in $\Gamma$ at MJD~56404.88. 

The improved \snr allows us to additionally let the folding energy $E_\text{fold}$, the energies of the two cyclotron lines, as well as energy of the \feka line vary. The results are shown in blue in Figure~\ref{fig:p2pres_all}, and follow the pulse-to-pulse results overall very well. Some ks-integrated results show significant deviations from the corresponding pulse-to-pulse results. These deviations originate in part from changes in parameters like  $E_\text{fold}$ and $E_\text{CRSF,H}$ which were frozen in the pulse-to-pulse analysis, but were allowed to vary in the ks-integrated data. Additionally, by summing over strong spectral changes on a pulse-to-pulse basis, the resulting ks-integrated spectrum might have an slightly different shape, resulting in a shifted best fit. These deviations, however, do not influence our conclusions.

The covering fraction CF is almost always larger than 0.8 and often consistent with 1, albeit with relatively large error bars in the pulse-to-pulse results. However, around MJD~56117.85 in observation I, the covering fraction drops significantly below 0.5, while at the same time \nh rises to very high values, on the order of $10^{24}$\,cm$^{-2}$. 
The photon index varies by about 0.3--0.4 in each observation, and is, as expected from the phase-averaged results, clearly different between observation I and II. 



Significant variations in the energy and depth of the harmonic CRSF are clearly visible in both observations (Figure~\ref{fig:p2pres_all}\textit{f} and \textit{g}).  The fundamental line strength changes from a clear detection in a single pulse to being below the detection limit, even in the ks-integrated data. In the last orbit of observation I the best-fit requires a strength $d_\text{CRSF,F}$ of 0, resulting in an unconstrained energy. We therefore fixed the line energy to half the energy of the harmonic line in order to estimate an upper limit of the line depth at this energy. In four other orbits, the line depth was consistent with 0 (90\% confidence), also leading to an unconstrained line energy. In these cases, we fix the strength to the best-fit value, and recalculate the energy confidence interval, shown in green in Figure~\ref{fig:p2pres_all}\textit{h}. This is the first time that the evolution of the cyclotron lines in \vela has been studied on these short time-scales.

The 10\,keV feature is most prominent at the beginning of observation I. Observation II almost never requires this feature for a good fit, either in pulse-to-pulse or in ks-integrated data. Only two spectra around MJD~56404.9 and MJD~56404.95 required a small but significant contribution from this feature. 
The energy of the 10\,keV-feature (not shown) is, when detected, in all cases very close to 10\,keV and does not show significant variation with time. 

To assess if the strong variability of the spectral parameters is independent of the chosen continuum, we also describe the ks-integrated spectra with the \texttt{NPEX} model. We obtain fits of similar quality in terms of \redchi. The parameters comparable between the models, i.e, the CRSF, the absorption and the iron line parameters are all consistent within the uncertainties and follow a similar evolution with time. We therefore rule out a systematic effect of the chosen continuum on the important spectral parameters.

\subsection{The spectrum of the off-state}
While the bright flare in observation I did not show any significant spectral changes, the off-state in observation II does (see Figure~\ref{fig:hrall}). We therefore extracted a spectrum covering only the off-state, which lasted for 450\,s before the hardness-ratio levels out and clear pulsations are visible in the lightcurve again. The statistics of this spectrum are comparable to the pulse-to-pulse spectra, but applying the same model gave a high value of $\chi^2=122.8$ for 90\,d.o.f., see Table~\ref{tab:bestfit_dip}. 

There is no evidence for a roll-over at higher energies, so we remove the 
\texttt{FDcut} component completely and describe the continuum with a power-law only. The best fit gives an acceptable  $\chi^2=104.2$ with also 90\,d.o.f.
The power-law index increases to a very high value, see the second column of Table~\ref{tab:bestfit_dip}. The absorption column density, however, does not rise to particularly high values, so the off-state is clearly not due to an absorption event. 

The cyclotron lines are not detected significantly in the off-state spectrum. The depth of the harmonic line is highly uncertain due to the lack of signal above 50\,keV. The overall spectral shape is very similar to the one found by \citet{doroshenko11a} in \suz data, though the \nustar data provide a significantly higher \snr above 30\,keV.

\begin{deluxetable}{rll}
\tablecaption{Fit parameters for the off-state.\label{tab:bestfit_dip}}
\tablewidth{0pc}
\tablehead{\colhead{Parameter}  & \colhead{FDcut} & \colhead{Powerlaw}}
 $ CF$ & $1.00^{+0.00}_{-0.21}$ & $1.00^{+0.00}_{-0.17}$ \\
 $ N_\text{H,1} [10^{22}\text{\,cm}^{-2}]$ & $11.6^{+6.8}_{-3.0}$ & $17^{+12}_{-4}$ \\
 $ \Gamma$ & $1.52\pm0.10$ & $1.91^{+0.14}_{-0.10}$ \\
 $ d_\text{CSRF,H} [\text{keV}]$ & $\le17$ & $20^{+20}_{-17}$ \\
 $ d_\text{CSRF,F} [\text{keV}]$ & $\le0.9$ & $\le1.7$ \\
 $ A(\text{Fe\,K}\alpha)^\dagger$ & $\left(0.9^{+1.0}_{-0.8}\right)\times10^{-3}$ & $\left(1.5^{+2.3}_{-1.5}\right)\times10^{-4}$ \\
 $ A(\text{Fe\,K}\beta)$\tablenotemark{$\dagger$} & $\le5\times10^{-4}$ & $\le1.2\times10^{-4}$ \\
 $ \mathcal{F}_\text{3--79\,keV}$\tablenotemark{$\star$} & $0.256\pm0.010$ & $0.319^{+0.025}_{-0.022}$ \\
 \tableline
 $\chi^2/\text{d.o.f.}$   & 122.82/90& 104.22/90\\$\chi^2_\text{red}$   & 1.365& 1.158
\enddata
\tablenotetext{$^\dagger$}{in ph\,s$^{-1}$\,cm$^{-2}$}
\tablenotetext{$^\star$}{in keV\,s$^{-1}$\,cm$^{-2}$}
\end{deluxetable}

\begin{figure}
\includegraphics[width=0.96\columnwidth]{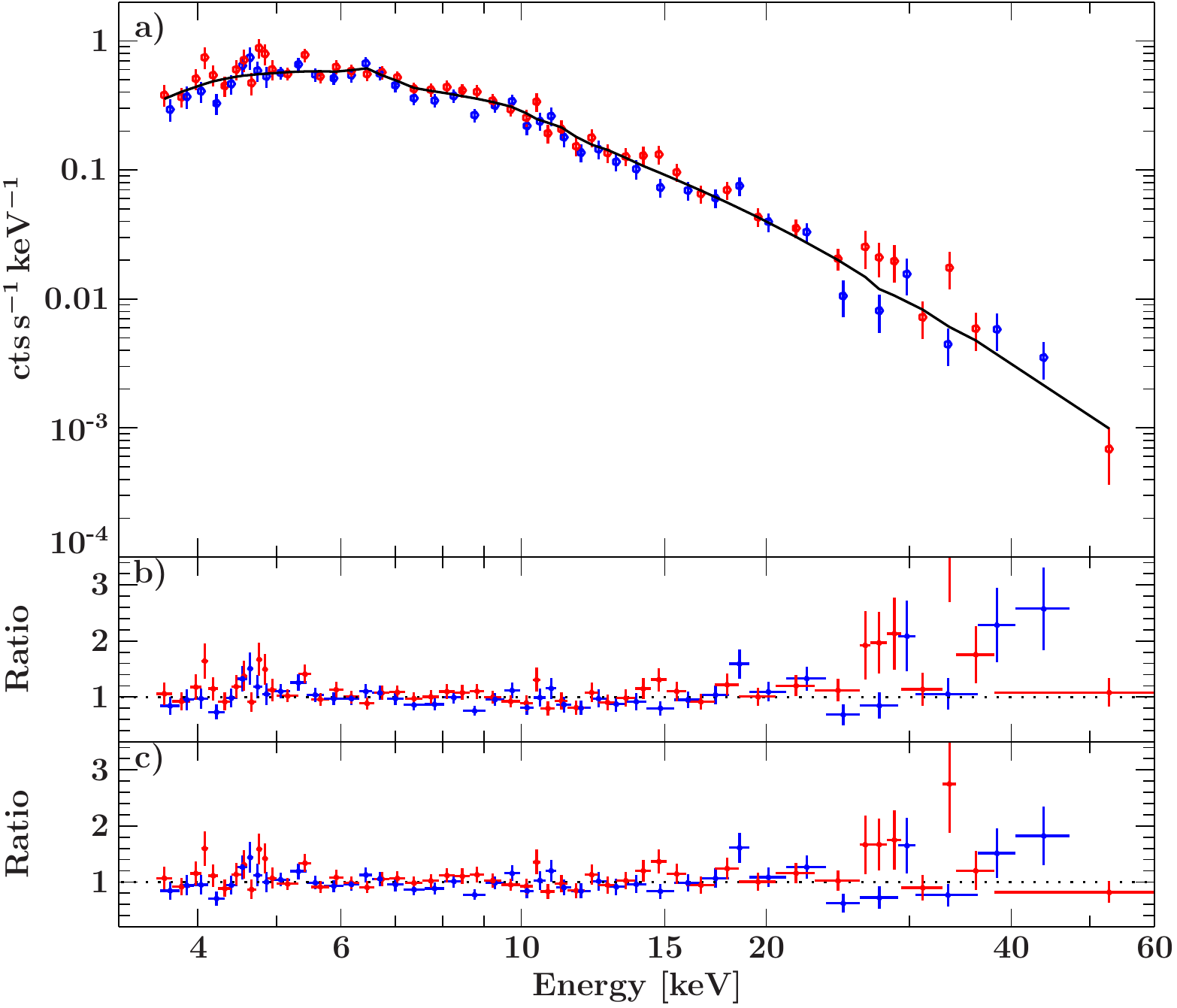} 
\caption{\textit{a)} Spectrum of the off-state with the best-fit power-law model in black, FPMA data in red and FPMB data in blue.  \textit{b)} Residuals with the standard pulse-to-pulse \texttt{FDcut} model \textit{c)} Residuals with power-law only continuum.}
\label{fig:specoff}
\end{figure}

\section{Discussion}
\label{sec:disc}

\subsection{Flux dependence of the cyclotron lines}
\label{susec:f2crsf}

\begin{figure}
\includegraphics[width=0.96\columnwidth]{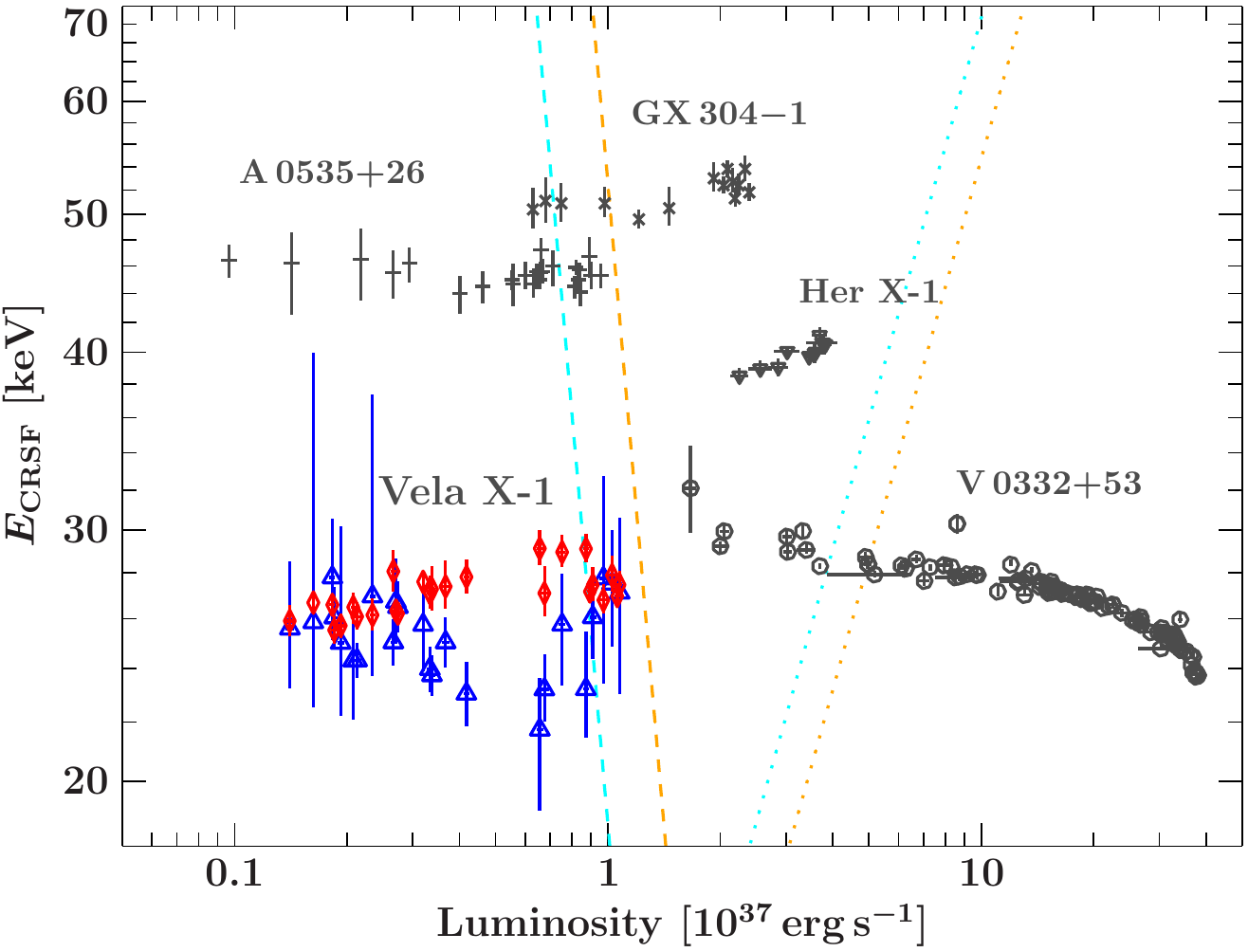} 
\caption{Fundamental cyclotron line energy as function of luminosity for different sources. 
The energy of the fundamental line of \vela is shown in blue using upright triangles, the harmonic line divided by a factor of two in red diamonds.
The dashed lines indicate the Coulomb breaking limit, $L_\text{Coul}$ and the dotted lines the critical luminosity, $L_\text{crit}$, following \citet{becker12a} and using $\Lambda=0.1$. The left-hand line (cyan) of each pair assumes $M_\star= 1.4\,\msun$, the right-hand line (golden) $M_\star=1.8\,\msun$. A\,0535+26 data (crosses) are obtained from \citet{caballero07a}, V\,0332+53 (circles) are from \citet{tsygankov10a}, Her~X-1 (down-pointing triangles) are from \citet{staubert07a}, and GX~304$-$1 (asterisks) are from \citet{yamamoto11a}.
}
\label{fig:lx2ecyc}
\end{figure}

Using the results from the ks-integrated spectra, we plot the energies of the CRSFs as function of luminosity in Figure~\ref{fig:lx2ecyc}.  A clear increase of the energy of the harmonic CRSF with luminosity is visible, which has not significantly been seen  in a persistent HMXB before. The evolution of the energy of the fundamental line with luminosity is difficult to interpret and influenced by the fact that the width is poorly constrained. Its energy  may also  be influenced by photon-spawning, as detailed in Sect.~\ref{susec:corrFH}.

To put the data into context, a sample of other sources and their CRSF-luminosity relation, as well as  $L_\text{coul}$ and $L_\text{crit}$ are also shown in Figure~\ref{fig:lx2ecyc}, following Figure~2 of \citet{becker12a}. The luminosity at which Coulomb interactions start to dominate, $L_\text{coul}$, and  the critical luminosity, where the local Eddington limit is exceeded, $L_\text{crit}$, are calculated under the same assumptions as in \citet{becker12a}. Additionally we show both luminosities for a neutron star with a higher mass of $M_\star=1.8\,\msun$, the lower mass limit of \vela. 

Recently \citet{hemphill13a}, using \inte data as well as values from the literature, showed that 4U~1907+09 shows an indication for a  correlation between line energy and luminosity at luminosities around $2\times10^{36}$\,erg\,s$^{-1}$. However, the correlation is strongly dependent on one measurement at high luminosities by \citet{rivers10a} and has systematic uncertainties due to comparing values from the literature measured with different continua and are therefore not shown in Figure~\ref{fig:lx2ecyc}.

From Figure~\ref{fig:lx2ecyc} it seems clear that \vela is situated in the regime where the in-falling matter is only stopped at the stellar surface, and no correlation between luminosity and line energy is expected. A\,0535+26 seems to be in the same regime and follows this predicted behavior. 
Nontheless the energy of the harmonic CRSF of \vela  shows a clear correlation with luminosity.
These data are the first clear evidence for a correlation below $L_\text{x}\sim10^{36}$\,erg\,s$^{-1}$.
In the following, we discuss a possible explanation for the clear correlation found in \vela at these low luminosities.


To calculate $L_\text{coul}$, at which the Coulomb interaction will stop the in-falling matter above the stellar surface in a shock, \citet{becker12a} first calculate the radius of the accretion column in their Eq. 23 with
\begin{align}
r = & 1.93\times \left(\frac{\Lambda}{0.1}\right)^{-\frac12}\left(\frac{M_\star}{1.4\,\msun}\right)^{-\frac{1}{14}}\left(\frac{R_\star}{10\,\text{km}}\right)^{\frac{11}{14}} \times \notag\\
& \times \left(\frac{B_\star}{10^{12}\,G}\right)^{-\frac{2}{7}}\left(\frac{L_x}{10^{37}\,\text{erg\,s}^{-1}}\right)^{-\frac{1}{7}}
\end{align}
where $M_\star$, $R_\star$, and $B_\star$, are the mass, radius and surface magnetic field of the neutron star, respectively. The parameter $\Lambda$ is defined by \citet{lamb73a}: $\Lambda=1$ for spherical accretion and $\Lambda < 1$ for disk accretion. That is, $\Lambda$ summarizes the interaction between the magnetic field and the surrounding medium, describing the Alfv\'en radius. It is the least well known variable in the equation and a likely way to explain why \vela is in a different accretion regime despite its low luminosity.

\vela accretes directly from the stellar wind, a scenario best described by $\Lambda=1$.
 \citet{becker12a} assume a stable accretion disk and set $\Lambda=0.1$ in their calculation. 
 Additionally $M_\star \geq 1.8$\,\msun for \vela. With theses values, we obtain $r\approx0.4$\,km for the observed luminosities of \vela in the spherical accretion case. This is a much smaller radius compared to $r\approx1.3$\,km for the standard disk accretion case.

The high pulsed fraction of \vela provides observational evidence for a very small accretion radius. If the X-rays are produced in a very localized region, the effects of the rotation of the neutron star will be more pronounced. Calculations taking relativistic light-bending into account are required to constrain the accretion column radius, but are beyond the scope of this paper. 

\begin{figure}
\includegraphics[width=0.96\columnwidth]{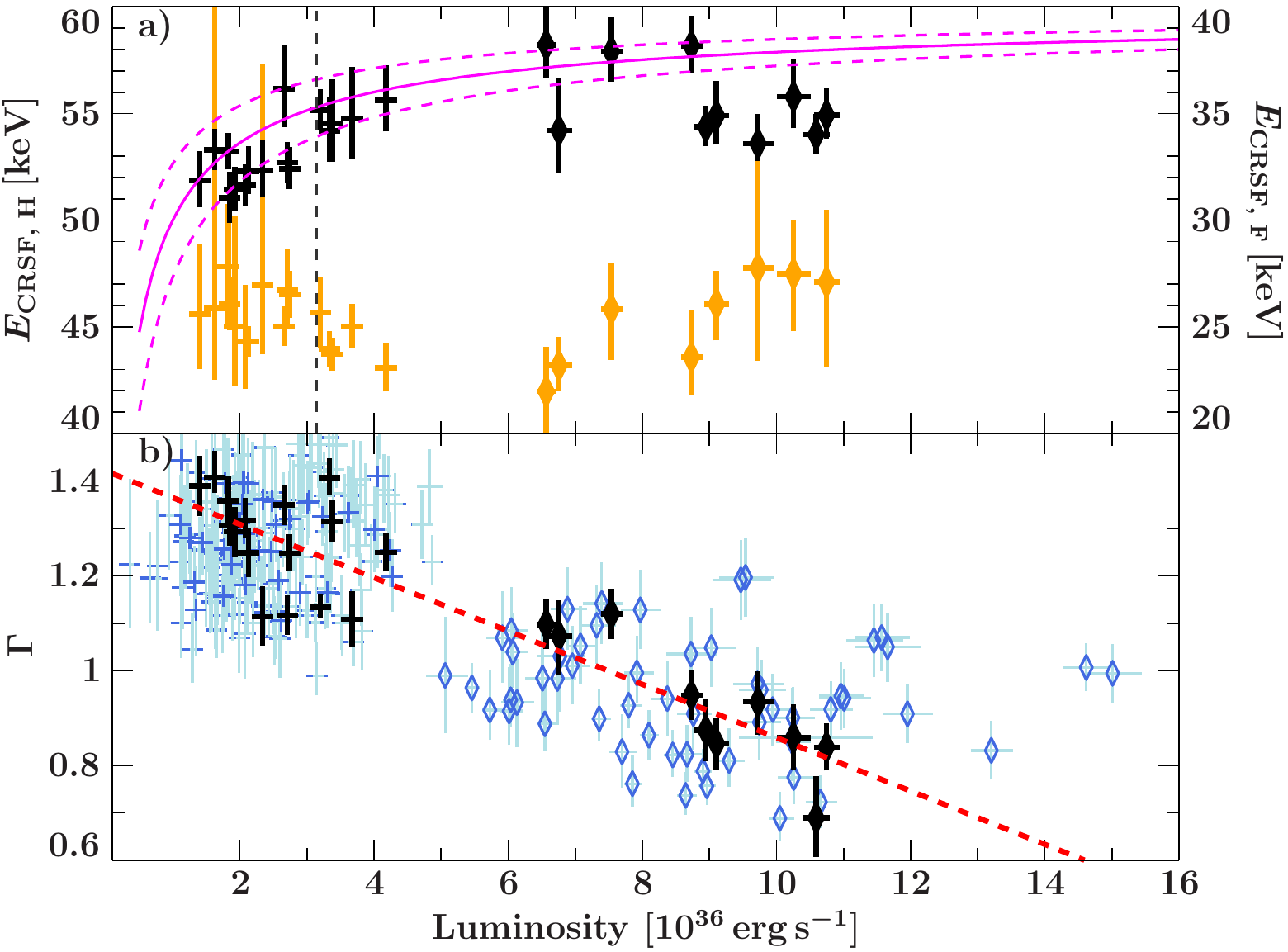} 
\caption{\textit{a)} Cyclotron line centroid energies as a function of luminosity between 3--79\,keV. Black points show the energy of the harmonic line,  orange points show  the energy of the fundamental line, using the right-hand $y$-axis. Diamonds and crosses indicate data taken during observation I and observation II, respectively. The magenta line is the theoretical prediction for $\Lambda=1$ and $M_\star=1.8\,\msun$, the dashed lines above and below it for $M_\star=1.4\,\msun$ and $M_\star=2.2\,\msun$, respectively.
As can be seen, the harmonic line energies follow the prediction well and are correlated with luminosity.
 The black dashed lines indicates $L_\text{coul}$ for our set of assumptions, using a surface magnetic field of $2.59\times10^{12}$\,G. For details see text. \textit{b)} Photon index $\Gamma$ as a function of X-ray luminosity. The red dashed line is a linear regression through all ks-integrated data points to guide the eye. Pulse-to-pulse data are shown in blue, ks-integrated in black.}
\label{fig:ge2flx}
\end{figure}

The drastically reduced radius of the accretion column directly leads to a reduced Coulomb luminosity. Assuming $B_\star= 2.59\times10^{12}$\,G, i.e., a surface fundamental cyclotron energy of $E_\star= 30$\,keV, we compute $L_\text{coul} = 3.34\times10^{36}$\,erg\,s$^{-1}$ using Eq.~54 of \citet{becker12a}. This luminosity is indicated by the vertical dashed line in Figure~\ref{fig:ge2flx}\textit{a}, moving most of our data above the limit.  
In this regime, the altitude of the line forming region decreases with luminosity, resulting in a higher line energy with higher flux, as observed.
The measured line energies also show a correlation with luminosity below this limit, but it is likely that Coulomb interactions start to decelerate the in-falling material already before the exact limit is reached.

 Using Eqs.~51 and 58 of \citet{becker12a}, the energy-luminosity correlation follows
\begin{align}
E_\text{theo} = &  \Biggl[1+ 0.6  \left(\frac{R_\star}{10\,\text{km}}\right)^{-\frac{13}{14}}  \left(\frac{\Lambda}{0.1}\right)^{-1} \left(\frac{\tau_\star}{20}\right)
\left(\frac{M_\star}{1.4\,\msun}\right)^{\frac{19}{14}} \times \notag\\
&\times  \left(\frac{E_\star}{1\,\text{keV}}\right)^{-\frac{4}{7}}\left(\frac{L_x}{10^{37}\,\text{erg\,s}^{-1}}\right)^{-\frac{5}{7}}\Biggr]^{-3}\times E_\star\quad\text{.}
\end{align}
Here $\tau_\star$ is the Thomson optical depth, estimated to be around 20 for typical HMXB parameters \citep{becker12a}.

This function is shown in magenta in Figure~\ref{fig:ge2flx}\textit{a}, using $\Lambda=1$ and $M_\star=1.8\,\msun$. It describes the data of the harmonic line qualitatively well up to about $9\times10^{36}$\,erg\,s$^{-1}$. At this luminosity the energy of the line drops to a relatively constant level around 54\,keV. The theoretically calculated energy is rather flat in this luminosity range, too, so that the discrepancy is possibly due to a constant offset, perhaps because $\Lambda$ is slightly different from 1.0. 

The mass of the neutron star has a strong influence on the correlation, and we show it for two other masses, $M_\star=1.4\,\msun$ and $M_\star=2.2\,\msun$, as dashed magenta lines in Figure~\ref{fig:ge2flx}\textit{a}. A lower mass moves the correlation to higher energies for a given flux and results in a poorer description of the data. A mass of $M_\star=2.2\,\msun$ provides the best description of the low luminosity data. Of course, the neutron star mass is only one of many parameters influencing this correlation, but the data seem to prefer a more massive neutron star.

\citet{odaka13a}, using \suz data taken in 2008, have shown that the photon index $\Gamma$ is anti-correlated with luminosity. This agrees well with the fact that the photon-index typically shows the opposite behavior as function of flux as the cyclotron line energy \citep{klochkov11a}. As shown in Figure~\ref{fig:ge2flx}\text{b}, our data show the same behavior, albeit with a much shallower slope.  However, \citet{odaka13a} used the \texttt{NPEX} continuum, which makes a direct comparison difficult. The hardening of the spectrum with X-ray flux can be understood as a direct consequence of the Coulomb shock moving further down into the accretion column where the electron gas becomes hotter. Photons can be scattered up to higher energies via the inverse Compton effect, and a harder spectrum is observed \citep{klochkov11a, becker07a}. The pulse-to-pulse results indicate that at the highest luminosities the relation flattens out, which is very similar to what we observed for the cyclotron line energies.

With the reduced accretion radius, the local Eddington limit is also reduced. Using Eq.~55 of \citet{becker12a}, we find that $L_\text{crit}$ is even below $L_\text{coul}$, i.e., \vela would be in the super-critical accretion regime. In this regime, an anti-correlation between the line energy and the luminosity would be expected, contrary to the behavior observed of the harmonic line. However, $L_\text{crit}$ has been calculated under the assumption that it can be scaled to the size of the accretion column compared to the surface of the neutron star. In the extreme case of a very narrow accretion column, this assumption might not hold, for example, radiation could escape through the walls of the accretion column before interacting with the in-falling material. Further calculations are needed to investigate this question.

Our model of sub-critical accretion with a narrow accretion column provides a good explanation of the measured behavior of the harmonic line. The fundamental line,  however, shows a very different behavior as a function of luminosity. In fact, there seems to be an anti-correlation visible up to 7$\times10^{36}$\,erg\,s$^{-1}$, at which point the correlation seems to flatten, i.e., it shows an opposite behavior as the harmonic line. 
It is difficult to conceive a physical reason for this discrepancy. 
Influence of the 10\,keV feature on the line energy is unlikely, as the behavior is also seen in observation II, in which the 10\,keV feature was not fitted in  the  ks-integrated spectra. 


\subsubsection{Accretion regime in the off-state}
We analyzed the spectrum of the off-state during observation II. We described it with an absorbed power-law, without the need for a exponential cut-off. While the photon-index of the power-law is higher in that state, the disappearance of the cut-off makes the continuum overall harder. We could not significantly detect a cyclotron line.
\citet{doroshenko13a} analyzed a similar off-state in \suz data and argue that the accretion cannot be switched off completely because  the spectrum is hard and residual pulsations remain visible.
The luminosity was $\left(2.45^{+0.19}_{-0.16}\right)\times10^{35}$\,erg\,s$^{-1}$ in the \nustar off-state data, i.e., about an order of magnitude lower than the lowest point shown in Figure~\ref{fig:ge2flx}. At that very low luminosity, it is unlikely that a shock is stopping the accretion flow, so the in-falling materials is only stopped at the surface of the neutron star. Here, the Comptonizing electron gas is hotter and denser, possibly moving the cut-off and the energy of the harmonic CRSF  above 60\,keV. This would be consistent with our findings of a single power-law component. 

\subsection{Correlation between the fundamental and the harmonic line}
\label{susec:corrFH}

\begin{figure}
\includegraphics[width=0.96\columnwidth]{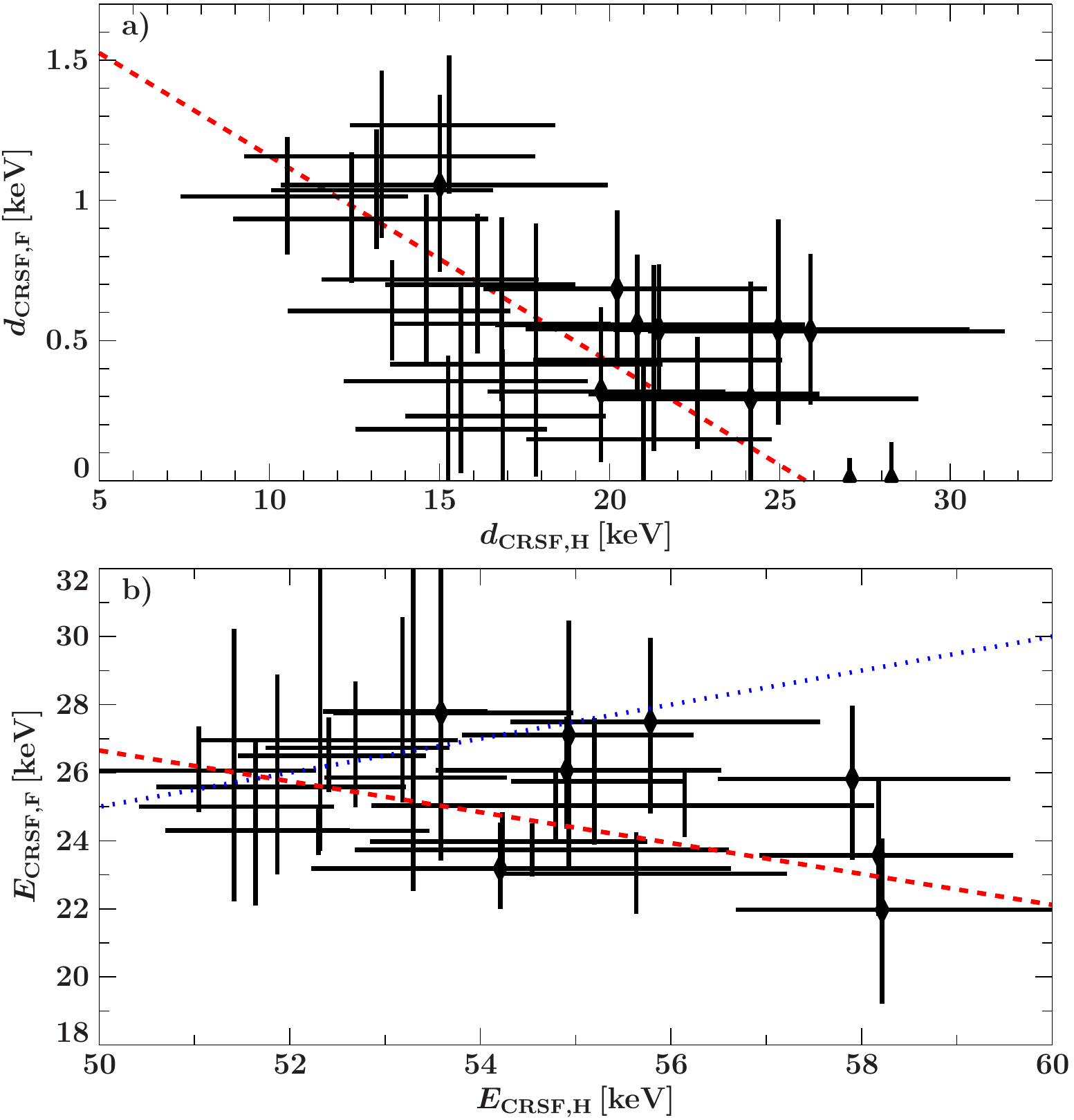} 
\caption{\textit{a)} Correlation between the depth of the harmonic CRSF $d_\text{CRSF,H}$ and the depth of the fundamental line $d_\text{CRSF,F}$. The red dashed line shows a linear fit to guide the eye. \textit{b)} Correlation between the line energies of the harmonic CRSF and the fundamental line. The red dashed line shows a linear fit to the data points, the blue dotted line indicates a factor of 2.0 between the line energies.}
\label{fig:crsfcorr}
\end{figure}

The \nustar data allow both cyclotron lines to be simultaneously measured and their evolution to be studied on time-scales as short as 1000\,s. As shown in Figure~\ref{fig:crsfcorr}\textit{a}, the depth of the fundamental and the harmonic line are anti-correlated, with a Pearson's correlation coefficient of $-0.68$. Spearman's rank correlation gives $\rho=-0.72$ with a false alarm probability of $3\times10^{-5}$, indicating a highly significant anti-correlation. Taking both $x$- and $y$-errors into account and using the routine described by \citet{williams10a} to fit a linear correlation, we obtain a best-fit slope of $-0.07\pm0.02$. As soon as the strength of the harmonic line $d_\text{CRSF,H}$ reaches $\sim$25\,keV (optical depth $\tau_\text{CRSF,H}\approx1.34$), the fundamental line drops below the noise limitations of the data.

This anti-correlation is likely a result of photon-spawning. Photon-spawning occurs when an electron at an excited Landau-level falls back to a lower state, emitting one or more photons close to the energy of the fundamental line energy \citep{schoenherr07a}. With an increasing depth of the harmonic line, more electrons are in excited levels, resulting in a higher number of spawned photons, which can fill up the fundamental line. Photon-spawning is a direct consequence from resonant scattering and explains the observed anti-correlation of the line depths.

The strong variability of the fundamental line depth can explain apparently contradictory results in the literature, concerning the presence of the line \citep[see, e.g., ][]{kretschmar97a, orlandini98a, kreykenbohm02a, labarbera03a, schanne07a, odaka13a}. 
For example, \citet{orlandini98a} did not find evidence for the fundamental line in \sax data. But these authors measured an equivalent width of the harmonic of $30\pm3$\,keV, much higher than our values (equivalent width of $\approx$8--14\,keV). This prominent line could provide enough photons to fill up the fundamental line such that it is undetected. On the other hand, the optical depth $\tau$ of the line in \xte data shown by \citet{kreykenbohm02a} is between 0.5--1, depending on pulse phase, close to our values of around $\tau=0.8$--0.9. In these data the fundamental line is clearly detected. 
We conclude that the fundamental CRSF is at 25\,keV, in agreement with \citet{kendziorra92a} but has avoided detection in some data due to photon-spawning from a prominent harmonic line. 

The ratio of the energies of the cyclotron lines is shown in Figure~\ref{fig:crsfcorr}\textit{b}. To first order, we would expect the line energies to be correlated, with a factor of two between the energy of the fundamental line and the harmonic line \citep{harding91a}.  The  dotted line illustrates this expected correlation. Our data clearly deviate from that correlation. When ignoring the  data in which the fundamental line was not detected, Pearson's correlation coefficient is $-0.35$ and Spearman's $\rho=-0.25$, with a false alarm probability of 0.23. The best fit slope of a linear fit is $-0.45\pm0.20$. This implies an anti-correlation with a little over $1\sigma$ confidence. Confidence contours for the time-averaged spectra show that there is no model intrinsic degeneracy present between the line energies.

The deviation from the canonical ratio of 2.0 is directly connected to the different behavior as function of X-ray luminosity, as shown in Figure~\ref{fig:ge2flx}. Spawned photons from the harmonic lines are likely to influence the shape of the fundamental line, and could probably shift the measured centroid energy from its real value. The data do not provide enough resolution to investigate the shape of the fundamental line in detail. This theory can therefore only be tested by extensive theoretical calculations, which can provide insight how the line energy ratio can deviate so strongly from 2.0.

\subsection{Variability and nature of the absorber}
The environment around the neutron star is dominated by the  heavily structured stellar wind from the optical companion.
This clumpy wind is responsible for the complex absorber, which we model as partial covering, with two independent absorption columns. Additionally a photo-ionization wake is present, which moves into our line-of-sight after orbital phase 0.5 \citep[see, e.g., ][]{blondin91a}, resulting in increased absorption at late orbital phases \citep[and references therein]{doroshenko13a}. We observe average column densities around $3\times10^{23}$\,cm$^{-2}$, which fit very well into that picture.

During observation I, \nh increases to a very high value of $\sim1.2\times10^{24}$\,cm$^{-2}$ (see Figure~\ref{fig:p2pres_all}\textit{c}). The column density then changes smoothly over about 10\,ks, decreasing again to values seen at the beginning of the observation. Assuming this event was caused by a clump moving through our line-of-sight we can calculate its size. At the orbit of the neutron star ($R = 1.7R_\star$), the wind has a speed of about $v_\text{wind}= 540$\,km\,s$^{-1}$ \citep{dupree80a}. At the orbital phase of the event, $\phi_\text{orb}=0.68$, the wind speed is almost perpendicular to the line-of-sight, so that if the clump moves with the same velocity, it stretches out for about $r_\text{cl} = 5.4\times10^{11}$\,cm. This is about an order of magnitude larger than the clumps calculated from X-ray flares by \citet{fuerst10a} and \citet{martinez13a}. We note that the wind speed might be drastically reduced due to the influence of the neutron star and its strong ionizing X-ray radiation \citep{krticka12a}. However, only very close to the neutron star does the wind velocity become low enough that a clump of the size of previous works ($r_\text{cl} \approx 10^{10}$\,cm) would stay in the line-of-sight for 10\,ks.

Following \citet{fuerst10a}, assuming a spherical clump with  $r_\text{cl} = 5.4\times10^{11}$\,cm and an over-density of 100 compared to the smooth wind ($\rho_\text{wind}= 8\times10^{-15}$\,g\,cm$^{-3}$)\footnote{Note that \citet{fuerst10a} give the wrong unit for $\rho$. The correct one would be kg\,cm$^{-3}$.}, results in an absorption column of $N_\text{H}^\text{cl}=2\times10^{23}$\,cm$^{-2}$. This is about a factor of 10 smaller than observed. 
From these estimates, it becomes clear that even an unusually large clump cannot explain the observations by simply moving through the line-of-sight with the wind speed.

There is some degeneracy between \nh and CF in the model. \nustar's effective area drops below 3.0\,keV so this degeneracy cannot be fully disentangled and the measured \nh values should be taken with some caution. The measured variations of \nh and CF are,  however, larger than the model degeneracy would allow, so that the increased absorption is very likely a real event.

It is possible that the event is caused not by one clump, but by a larger number of them, moving in such a way that they start to overlap during the phase of increased absorption.  According to stellar wind theories, the vast majority of the wind is located in clumps, so that many of them are close to our line-of-sight at any given moment \citep[see, e.g., ][]{oskinova07a}. When the clumps start to overlap, the absorption column increases and at the same time the covering fraction can decrease, as observed.
The assumption of spherical clumps, however, is not necessarily correct. Simulations by \citet{dessart05a}, for example, show that line-driven stellar winds show complex shocked structures and not simple spherical clumps. Elongated clumps aligning with the line-of-sight could explain the observed behavior without  the need of very large size. Detailed simulations would, however, be necessary to quantify this scenario.

\section{Summary \& Outlook}
\label{sec:summ}
Using two different \nustar observations, we performed spectral analysis with high temporal and spectral resolution. The average spectrum can be well described with a standard absorbed \texttt{FDcut} model. 
The \nustar data have unequivocally shown that \vela shows two cyclotron lines, with the fundamental line at about 25\,keV.  By performing  spectral analysis on pulse-to-pulse and kilo-second time-scales, we have shown that the absorption, continuum and cyclotron line parameters are all highly variable.
We showed that the strength of the fundamental line is anti-correlated with the strength of the harmonic line. This effect can be explained by photon-spawning \citep{schoenherr07a}. 

For the first time, we have measured a correlation of the cyclotron line energy with luminosity. Using a surface magnetic field of $2.6\times10^{12}$\,G, we can describe this correlation with the theoretical predictions of \citet{becker12a}. Assuming sub-critical accretion with a Coulomb-dominated shock and a narrow accretion column, the data prefer a neutron star mass in excess of 1.4\,\msun. The mass is consistent with an independent and precise study of the optical companion and the orbit \citep{quaintrell03a}.

Our collaborations is working on new models which will become available soon and describe the production of hard X-rays in the accretion column based on first principle calculations, combined with the formation of cyclotron lines and the influence of the strong magnetic field on the scattering cross-sections. Applying these models to \vela will allow for a description of the physical conditions inside the accretion column in more detail, and prove or disprove that \vela has a very narrow accretion column, accreting in the sub-critical regime.

\acknowledgments
This work was supported under NASA Contract No. NNG08FD60C, and
made use of data from the {\it NuSTAR} mission, a project led by
the California Institute of Technology, managed by the Jet Propulsion
Laboratory, and funded by the National Aeronautics and Space
Administration. We thank the {\it NuSTAR} Operations, Software and
Calibration teams for support with the execution and analysis of
these observations. This research has made use of the {\it NuSTAR}
Data Analysis Software (NuSTARDAS) jointly developed by the ASI
Science Data Center (ASDC, Italy) and the California Institute of
Technology (USA). We would like to thank John E. Davis for the \texttt{slxfig} module, which was used to produce all figures in this work. 
We would like to thank Fritz Schwarm for the helpful discussions about cyclotron line shapes.
JAT acknowledges partial support from NASA Astrophysics Data Analysis Program grant NNX13AE98G. MB was supported by the Centre National d'\'Etudes Spatiales (CNES). We would like to thank the anonymous referee for the useful comments.

{\it Facilites:} \facility{NuSTAR}

%



\end{document}